\documentclass[showpacs,showkeys,preprint,prd,nofootinbib, 12pt]{revtex4-1}

\usepackage{graphicx} 

\usepackage{amsmath}
\usepackage{subfigure}
\usepackage{epstopdf}
\usepackage{multirow}
\usepackage{xspace}
\usepackage{rotating}
\usepackage{longtable}
\usepackage{multirow}
\usepackage{cancel}
\usepackage[breaklinks=true]{hyperref}
\hypersetup{
  colorlinks=true,
  linkcolor=blue,
  citecolor=blue,
  urlcolor=blue
}

\makeatletter
\renewcommand*\env@matrix[1][*\c@MaxMatrixCols c]{%
  \hskip -\arraycolsep
  \let\@ifnextchar\new@ifnextchar
  \array{#1}}
\makeatother

\graphicspath{{./fig/}}
\usepackage{array,tabularx,epsfig,mathrsfs,graphicx,rotating}
\usepackage{ifthen}
\usepackage{amsfonts}
\newcommand{\beq}{\begin{equation}}
\newcommand{\eeq}{\end{equation}}

\chardef\til=126

\newcommand{\vect}[1]{\boldsymbol{#1}}
\newcommand{\etmiss}{E_T^{\mathrm{miss}}}

\begin{document}

\preprint{ANL-HEP-144006}

\hfill{version 3, March 13, 2019}


\vspace{2.5cm}

\title{
Imaging particle collision data for event classification \\ using 
machine learning  
}

\author{S.~V.~Chekanov}
\affiliation{
HEP Division, Argonne National Laboratory,
9700 S.~Cass Avenue, Argonne, IL 60439, USA 
}%

\begin{abstract}
We propose a method 
to organize  experimental data from particle collision experiments 
in a general format which can enable a simple visualisation and effective classification 
of collision data using machine learning techniques.
The method is based on sparse fixed-size matrices  
with single- and two-particle variables 
containing information on identified  particles and jets. 
We illustrate this method using an example of searches for new physics at the LHC experiments.
\end{abstract}



\maketitle


\section{Introduction}

Machine learning is successfully used 
for classification of experimental data in particle
collision experiments (for a review see \cite{1742-6596-608-1-012058}). 
This technique becomes increasingly important in searches for  
new physics in billions of events collected by the LHC experiments.
Studies of interesting physics channels using machine learning  
usually includes an identification of most relevant  (``influential'') input variables, data reduction, 
data re-scale (the range  [0, 1] is a popular choice),
dimensionality reduction,   
data normalization (to avoid cases when some input values overweight others) and so on. 

Among various supervised-learning techniques,
Neural networks (NN) \cite{Bishop:2006,Guest:2018yhq} are 
widely used for high-accuracy image 
identification and classification. 
In the case of large volumes of  multi-dimensional data, 
such as data with information on the final-state particles and jets produced by particle accelerators, 
the usage of the NN is more challenging.   
A procedure should be established to find influential input variables 
from the variable-size lists with characteristics of particles and jets.
Examples of analysis-specific input variables for supervised  machine learning  
designed for reconstruction of heavy particles produced in $e^+e^-$ and $pp$  experiments   
can be found in \cite{Chekanov:2003cp,Santos:2016kno}.

The usage of machine learning in particle physics can be simplified 
if experimental data are transformed into image-like
data structures that can capture most important kinematic event characteristics. 
In this case, an identification of influential variables 
for specific physics signatures and further preparation of such variables 
(rescaling, normalization, decorrelation, etc.) for machine learning can be minimized.
At the same time, one can leverage  a wide range of algorithms 
for image classification developed by leading industries.
Similar to pixelated images of jets \cite{Cogan2015} for 
machine learning (for a review see \cite{Guest:2018yhq}),
a pixelated representation of kinematics of particles and jets produced in combination with
machine learning techniques  may shed light
on new phenomena in particle collisions. 

This paper proposes a mapping of particle records from colliding experiments to 
matrices that cover a wide range of properties of the final state.
Such 2D arrays will have pre-defined fixed sizes and fixed ranges of their values, 
unlike the original data records
that have varying number of particles represented by four-momenta or other (typically,  unbounded) 
kinematic variables.
In this context, the word ``imaging'' used in this paper refers to a pixelated representation
of kinematic characteristics of particles and jets in the form of such matrices.
We will show that these matrices can easily be visualized, and can
be conveniently interpreted by popular machine learning algorithms.

\section{Rapidity-mass matrix (RMM)}

Imaging event records with final-state particles means transforming
such data into a fixed-size grid of values in a given range.
In the simplest case, this can be a 2D matrix comprised of columns and rows 
of values that carry useful features of events to be used in event classification.
We propose to  construct a square matrix  of
a fixed width $1+\sum_{i=1}^T  N_{i}$, 
where $T$ is the total number of object types (jets, identified particles, etc.)
to be considered and $N_i$ is the expected maximum multiplicity
of an object $i$ in all events.
The  values $T$ and $N_i$ should be defined based on expectations for the types of reconstructed particles and
on technical capabilities of
the available computational resources for machine learning.
This matrix will contain values of single- and double-particle 
characteristics/properties of all reconstructed  objects. 

To be more specific, let us consider a dimensionless  matrix with $T=2$.     
The two objects to be considered are jets ($j$) and muons ($\mu$). 
We assume that the maximum number for each particle types is fixed to a constant $N$, i.e. $N_i=N$ for $i=1,2$. 
Then we define the following rapidity-mass matrix (RMM) for a given $pp$ collision as: 
 
\begin{equation}
\begin{pmatrix}[rrrrrrrr]
     \bf{e_T^{miss}}  &   m_T(j_1)     &  m_T(j_2)           &  \dots m_T(j_N)          &   m_T(\mu_1)    &  m_T(\mu_2)   &  \dots m_T(\mu_N)       \\
    h_L(j_1)   &   \bf{e_T(j_1)}       &  m(j_1,j_2)         &  \dots m(j_1,j_N)        &   m(j_1,\mu_1)  &  m(j_1,\mu_2) &  \dots  m(j_1,\mu_N)       \\ 
    h_L(j_2)   &  h(j_1,j_2)    &  \bf{\delta e_T(j_2)}        &  \dots m(j_2,j_N)        &   m(j_2,\mu_1)   &  m(j_2,\mu_2)&  \dots  m(j_2,\mu_N)     \\ 
   \dots    &   \dots             &  \dots              &  \dots,                  &   \dots          &  \dots       &  \dots                   \\ 
   h_L(j_N)   &  h(j_1,j_N)   &  \dots              &  \dots \bf{\delta e_T(j_N)}       &   m(j_N,\mu_1)   &   m(j_N,\mu_2) &  \dots  m(j_N,\mu_N)        \\ 
   h_L(\mu_1)  &  h(\mu_1,j_1) &  h(\mu_1,j_2)  &  \dots h(\mu_1,j_N) &   \bf{e_T(\mu_1)}  &   m(\mu_1,\mu_2)   &  m(\mu_1,\mu_N)        \\ 
   h_L(\mu_2)  &  h(\mu_2,j_1) &  h(\mu_1,j_2)  &  \dots h(\mu_2,j_N) &   h(\mu_1,\mu_2)  &  \bf{\delta e_T(\mu_2)}  &  m(\mu_2,\mu_N)        \\ 
   \dots      &   \dots        &  \dots         &  \dots                   &    \dots            &  \dots       \\ 
   h_L(\mu_N)  &   h(\mu_N,j_1) &  h(\mu_N,j_2)  &  \dots h(\mu_N,j_N) &    h(\mu_N,\mu_1)   &  h(\mu_N,\mu_2)  &  \bf{\delta e_T(\mu_N)}        \\ 
\end{pmatrix}
\label{eq1}
\end{equation}

\vspace{0.7cm}

The first element at the position (1,1) (shown in bold)  contains an event   
missing transverse energy $\etmiss$ 
scaled by $1/\sqrt{s}$, where $\sqrt{s}$ is a center-of-mass collision energy,
i.e. $e_T^{miss}=\etmiss /\sqrt{s}$.
The missing transverse energy $\etmiss$  is the magnitude
of missing transverse momentum vector $\vect{E}_{T}^{\mathrm{miss}}$ defined as the projection  
of the negative vector sum of all the reconstructed particle momenta onto the plane
perpendicular to the direction of colliding beams. 
Other diagonal cells contain the ratio  $e_T(i_1)=E_T(i_1)/\sqrt{s}$, where
$E_T(i_1)$ is the transverse energy of a leading in $E_T$ object $i$ (a jet or $\mu$),  and
transverse energy imbalances 
$$
\delta e_T(i_n) = \frac{E_T(i_{n-1})-E_T(i_{n})}{E_T(i_{n-1})+E_T(i_{n})}, \quad n=2,\ldots, N,  
$$ 
for a  given object type $i$.
All objects inside the RMM  are strictly ordered in transverse energy i.e.  $E_T(i_{n-1})>E_T(i_{n})$.
Therefore, $\delta e_T(i_n)$ always have positive values.  
The non-diagonal upper-right values are $m(i_n,j_k)=M_{i,n,\> j,k}/\sqrt{s}$, where $M_{i,n,\> j,k}$
are two-particle invariant masses.
The  first row contains transverse masses $M_T(i_n)$ of objects $i_n$ 
for two-body decays with undetected particles, 
scaled by $1/\sqrt{s}$, i.e.  $m_T(i_n)=M_T(i_n)/\sqrt{s}$.
The transverse mass $M_T$ is 
defined\footnote{For massless particles, $M_T$ can be approximated with
$\sqrt{2E_T \etmiss  (1-\cos(\Delta \phi))}$,
where $\Delta \phi$ is the opening angle between $\vect{p_T}$ and $\vect{E}_{T}^{\mathrm{miss}}$. 
We use the massless approximation in this paper.}
using the missing transverse momentum vector  $\vect{E}_{T}^{\mathrm{miss}}$, transverse energy $E_T$  and 
transverse momentum vector $\vect{p_T}$ of the observed particle/jet at the position $i_n$  as  
$M_T=\sqrt{ (E_T + \etmiss)^2 - (\vect{p_T}+ \vect{E}_{T}^{\mathrm{miss}})^2}$.     
According to this definition, $m_T(i_n)=0$ for $e_T^{miss}=0$ for massless particles. 

The first column vector is $h_L(i_n) = C (\cosh(y)-1)$, where $y$ is the rapidity of a 
particle $i_n$, and $C$ is a constant.
The rapidity of a particle/jet is defined in terms its energy momentum components $E$ and $p_z$ as 
as $y=0.5 \ln ((E+p_z) / (E-p_z))$. 
The variable $h_L(i_n)$ is proportional to the Lorentz factor, $\cosh(y)$,
thus it reflects longitudinal directions.  
The scaling factor $C$ is defined such that the average values of $h_L(i_n)$ are similar to 
those of $m(i_n,j_k)$ and $m_T(i)$, which is important for
certain algorithms that require input values to have similar weights.
This constant, which is found to be 0.15 
from Monte Carlo simulation studies for QCD multijet events, is sufficient to ensure similar orders 
of the magnitude for average values of  $h_L(i_n)$ and the scaled masses.  
The value 0.15 is also sufficient 
to make sure that $h_L(i_n)$ belongs to the interval
[0, 1] for a typical rapidity 
range $[-2.5, 2.5]$ for reconstructed particles and jets for collider experiments.
For other experiments with a different rapidity range, the constant $C$ needs to be recalculated.

The value $h(i_n,j_k)=C( \cosh(\Delta y / 2  )-1 )$ is constructed from the 
rapidity differences $\Delta y=y_{i_k} - y_{j_n}$  between $i$ and $j$.
The transformation with the $\cosh()$ function is needed to rescale  $\Delta y$ 
to the range $[0, 1]$ used by other variables.
For convenience, we will drop the indices $n$ and $k$ in later discussion.
 
Let us consider the properties of the RMM. All its values are dimensionless, and most variables are Lorentz 
invariant under boosts 
along the longitudinal axis (except for $h_L$ which itself defines the Lorentz factors). 
All RMM  values vary in the range $[0, 1]$ as 
required by many NN algorithms that transpose the input variables into the data range 
of sigmoid-like activation functions. This simplifies the usage of machine learning algorithms since the input 
rescaling to a fixed range of values is not needed.
The number of cells with non-zero values reflects 
the  most essential characteristics of collision events - multiplicities of reconstructed objects above 
certain kinematic requirements.
The number of reconstructed objects in an event can be obtained by 
summing up cells with non-zero values in rows (columns), and subtracting 1.
The  RMM matrix is very sparse for collision events with low multiplicities of identified  particles.

There is one important aspect of the RMM: the  vast majority of its cells have near-zero correlation 
with each other. 
Thus, to a large extent, the RMM does not contain redundant information.
Trivial effects, such as energy and momentum conservation, do not significantly contribute to the RMM.  
This can be seen from the structure of this matrix: rapidities and invariant masses
are expected to be independent.  
As a check, we calculated
the Pearson correlation coefficients, $\rho$, for all pairs of cells using the Monte Carlo event samples 
generated as discussed in Sect.~\ref{sec:vis}. 
The total number of cell pairs with non-zero values was found to be 1282.
Out of this number,
correlation coefficients with $|\rho|>0.5$ were observed for $5\%$ of cell pairs. They were typically
related to the cells with $m_T(i)$ and $e_T(i)$ variables.  
The number of cell pairs with $|\rho|>0.1$ was $15\%$.
The pairs of cells with Pearson correlation coefficients values larger than $15\%$  are given 
in the Appendix~\ref{app:corr}. 

Let us discuss the properties of the RMM relevant to physics signatures in particle experiments:

\begin{itemize}

\item
The first cell at the position (1,1) contains $\etmiss$ which is a crucial characteristic of events
in many physics analyses at the LHC. This variable is important for searches for new physics in
events with undetected particles,
but also for reconstruction of Standard Model particles. 

\item
The first row of values are also sensitive to 
missing transverse energy. They reflect  
masses of particles that include decays to invisible particles.
The most popular example is the $W\to l\nu$ decay for which the $m_T(i)$ cells carry information on the $W$ mass.
Transverse masses are used  in many searches  to  separate  the  signal  from  backgrounds since
they contain information on correlations of $\etmiss$ with other objects in an event.
For example, searches for SUSY (for example, see \cite{Tovey:2008ui} and references therein) 
and dark matter particles (for a review see \cite{Boveia:2018yeb}) will 
benefit from analysis of the first row.

\item
The first column of the RMM reflects longitudinal characteristics of events. It can be used for separation
of forward production from centrally produced objects.
For example, if jets are produced preferentially in the forward region, then $h_L(i)$ have non-zero values.  
This is  important in identification of events with hadronic activity in the forward region. For example,
the production of the Higgs ($H$) boson in the Vector Boson Fusion mechanism has usually at least one jet in 
the forward direction \cite{1742-6596-934-1-012030}.

\item
The diagonal elements  with $e_T(i)$ and $\delta e_T(i)$ values can be used for calculations
of transverse energies of all objects and, therefore, the total transverse energy of events.
The transverse energy imbalance, $\delta e_T$,  is sensitive 
to interactions of partons in the medium of heavy ion 
collisions (see, for example, \cite{PhysRevC.84.024906}). 
They can also be used  for a  separation  of multi-jet QCD production from more complex processes. 
Note that the energy $E(i)$ of an object $i$ can be reconstructed from $e_T(i)$ and $h_L(i)$ as 
$E(i) = e_T(i) \sqrt{s} (h_L(i) / C + 1)$. 
 
\item
The non-diagonal top-right cells 
capture two-particle invariant masses.  
For two-particle decays,  $m(i,j)$ are proportional to the masses of decaying particles. 
For example, for a resonance production, such as $Z$-bosons decaying to muons,  cells at 
$m(\mu_1,\mu_2)$  will be filled with the nominal mass of the $Z$ boson (scaled by $1/\sqrt{s}$).

\item
The RMM contains the information on rapidity differences via $h(i,j)$ (the lower left part of the RMM).
Collimated particles will have values of $h(i,j)$ close to zero. 
The rapidity difference between jets is often  used  in searches for 
heavy resonances \cite{Aaboud:2017yvp},  and is sensitive to
parton dynamics beyond collinear factorization \cite{Chatrchyan:2012pb}.   

\end{itemize}

The matrix Eq.~\ref{eq1} does not contain the complete information on four-momentum of each particle
(or other kinematic variables, such as the azimuthal angle $\phi$). 
In many cases, additional single-particle kinematic variables, such as $\phi$,  
are featureless due to the  rotational symmetry around the beam direction. 
Nevertheless, the RMMs themselves  can be used  for object selection and a basic data analysis: 
when interesting candidate events are identified using RMM,
one can refine the search using the RMM itself.
For example, in the case of searches for new resonances decaying to two leading jets,
one can  sum up the RMM cells at the position (3,2) in order to obtain distributions with 
two-jet invariant masses.  
Such an event-by-event analysis of RMMs may require a smaller data volume compared to the complete
information on each produced particle, since there are many techniques for 
effective storage of sparse matrices.  

The  matrix Eq.~\ref{eq1} can be extended to electrons, photons, $b-$tagged jets and reconstructed $\tau$'s. 
The RMM can also
be generalized to a 3D space after adding
three-particle kinematic  variables, such as three-particle invariant masses.


\section{Visualising collision events}
\label{sec:vis}
 
Let us give an example of how the RMM
approach can be used for visualisation and identification of different event types.
We will construct an RMM with four types ($T=4$) of reconstructed objects:  
jets ($j$), muons ($\mu$), electrons ($e$) and photons ($\gamma$). Three  
particles per type ($N_n=3$, $n=1,\ldots,4$) will considered.
This leads to a matrix  (to be denoted as T4N3) of the size of $13\times 13$ (one extra row and column contain $m_T(i)$ and $h_L(i)$).

We used the Pythia8 \cite{Sjostrand:2006za,Sjostrand:2007gs} generator 
for several Standard Model processes at a $pp$ collider  at $\sqrt{s}=13$~TeV.
Pythia8 uses the NNPDF 2.3 LO \cite{Ball:2012cx, Ball:2014uwa} parton density function
from the LHAPDF library \cite{Buckley:2014ana}.
We generated 50,000 Monte Carlo events at leading-order QCD for three processes:
multijet QCD, Standard Model Higgs production and top ($t\bar{t}$) production.
A minimum value of 200~GeV on the invariant mass of the $2\to2$ system was set during the event generation.
All decay channels of top quarks, $H$ and  vector bosons were allowed. 
This represents a particular difficult case since there are no unique decay signatures.

\begin{figure}[t]
\begin{center}
   \subfigure[RMM for hadronic decay of $t\bar{t}$] {
   \includegraphics[width=0.47\textwidth]{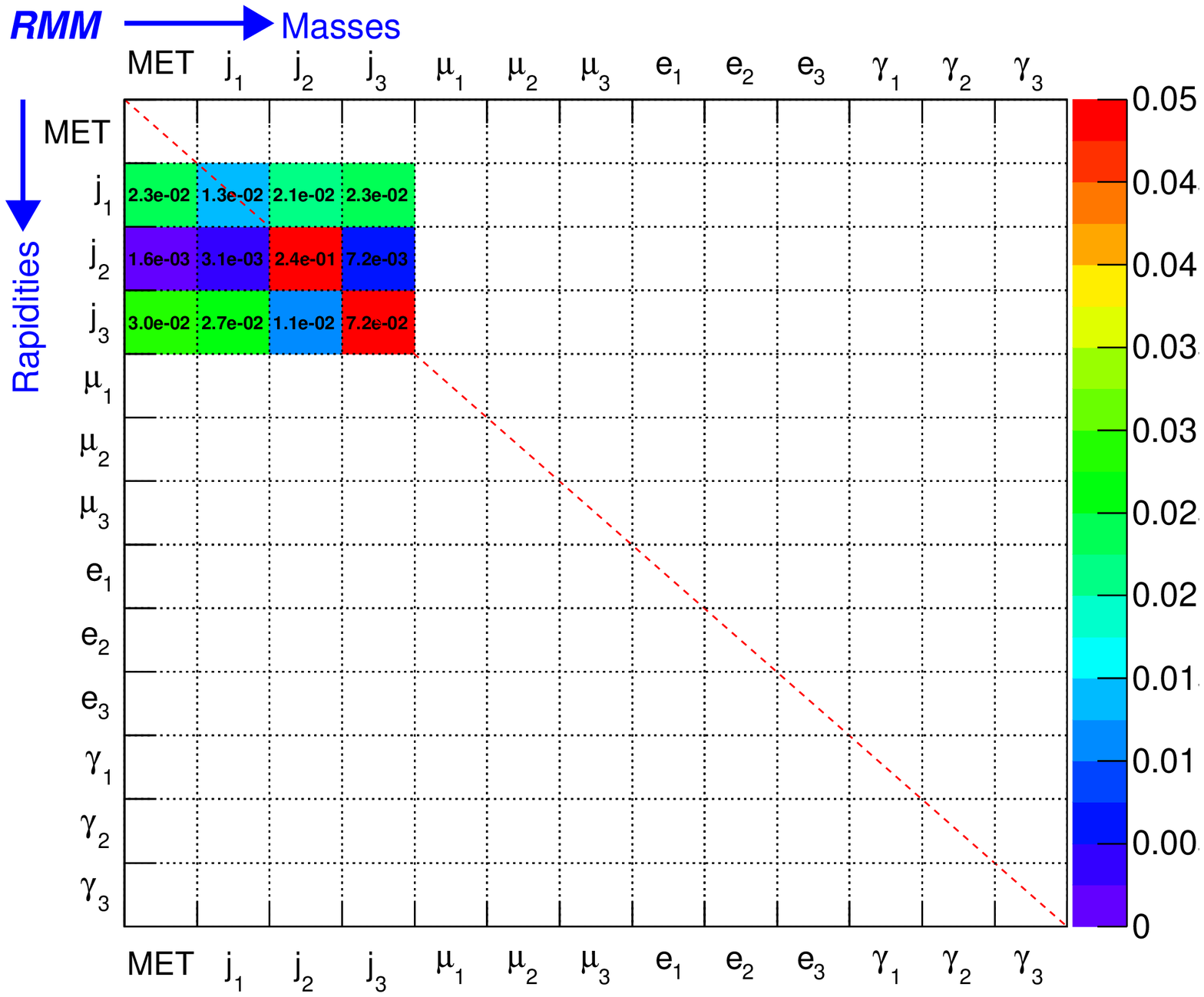}\hfill
   }
   \subfigure[RMM for semileptonic decay of $t\bar{t}$] {
   \includegraphics[width=0.47\textwidth]{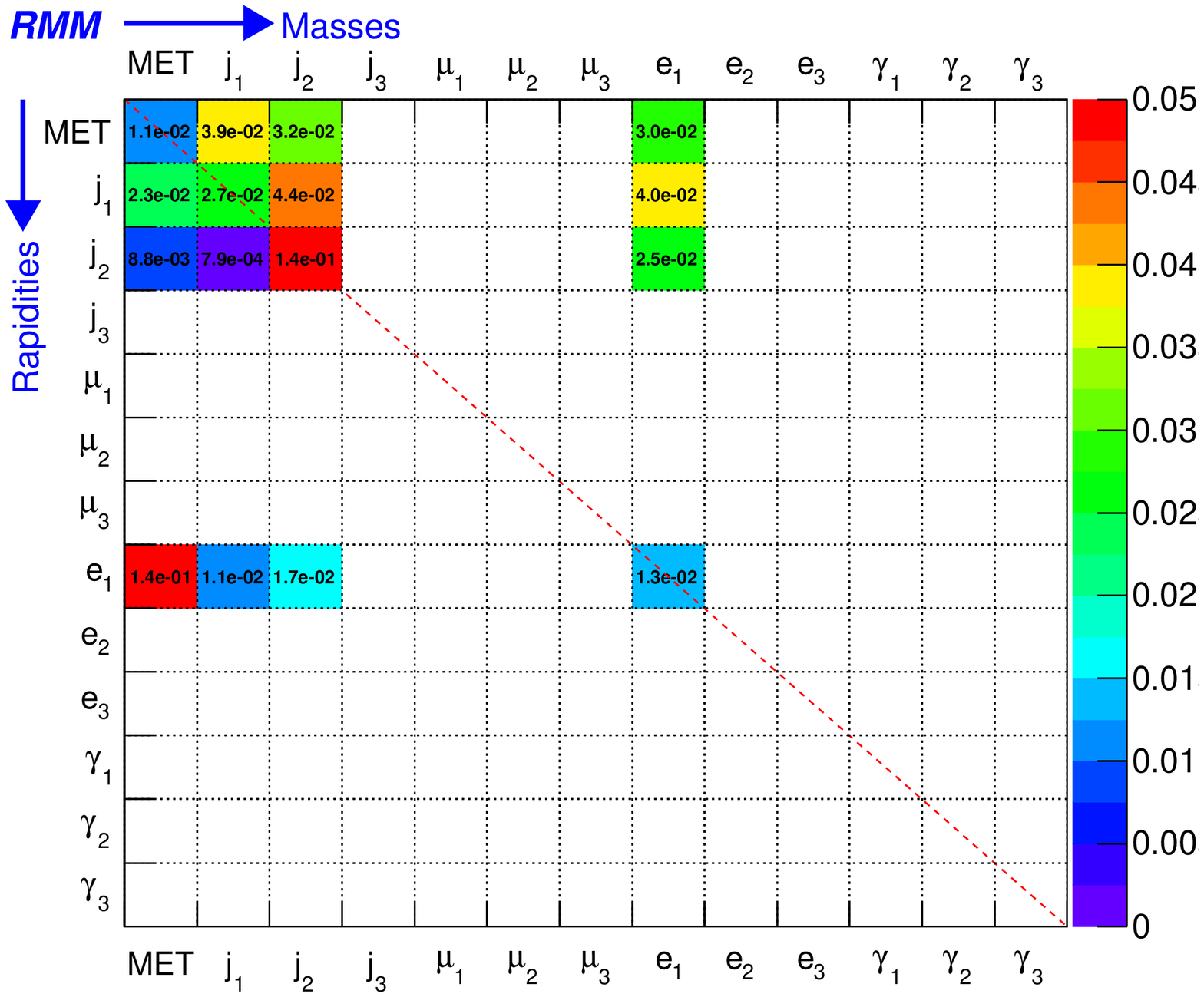}
   }
\end{center}
\caption{Visualization of the T4N3 RMMs for two $t\bar{t}$ events: (a) an event where $t$($\bar{t}$) decay
to six jets. (b) an event where one top quark decays to $\bar{b}W^+$ with $W^+\to e^+\nu_e$.
The latter event activates the cell at the  position (1,1) which
contains the information on $\etmiss$  scaled by $1/\sqrt{s}$.
The first row contains scaled transverse masses, $m_T(i)$.
The red dashed line shows the diagonal elements with transverse
energy and their imbalances. The upper-right part of this matrix contains $m(i,j)$, while the lower-left
part contains rapidity differences, $h(i,j)$.
The type of objects for each RMM cell is indicated.}
\label{fig:event}
\end{figure}

In addition to the Standard Model processes, 
events with charged Higgs boson ($H^+$) process were generated using  the diagram  $bg\to H^+t$, which is 
an attractive  exotic process  \cite{Akeroyd:2016ymd} arising  in 
models with two (or more) Higgs doublets.
This process was also simulated with Pythia8 assuming a mass of 600~GeV for the $H^+$ boson, which  
decays to $W^+$ and $H$.
In order make the identification of this process more challenging for our later discussion, 
we will consider $H$ decaying to two $b$-jets.
In this case, the event signatures (and the RMM values) are  rather similar to those from 
the $t\bar{t}$ production.

\begin{figure}[t]
\begin{center}
   \subfigure[Multi-jets QCD] {
   \includegraphics[width=0.4\textwidth]{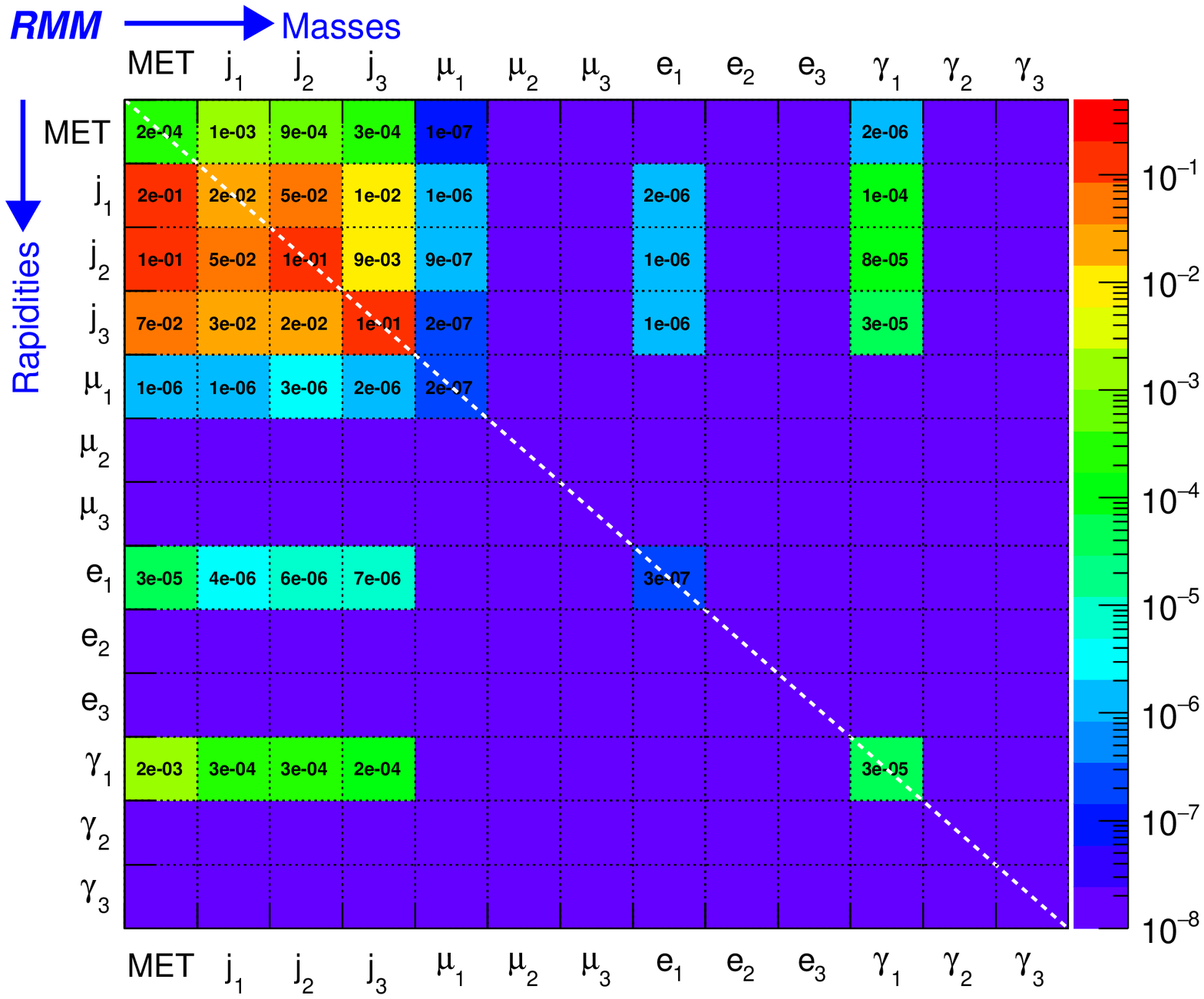}\hfill
   }
   \subfigure[Higgs processes] {
   \includegraphics[width=0.4\textwidth]{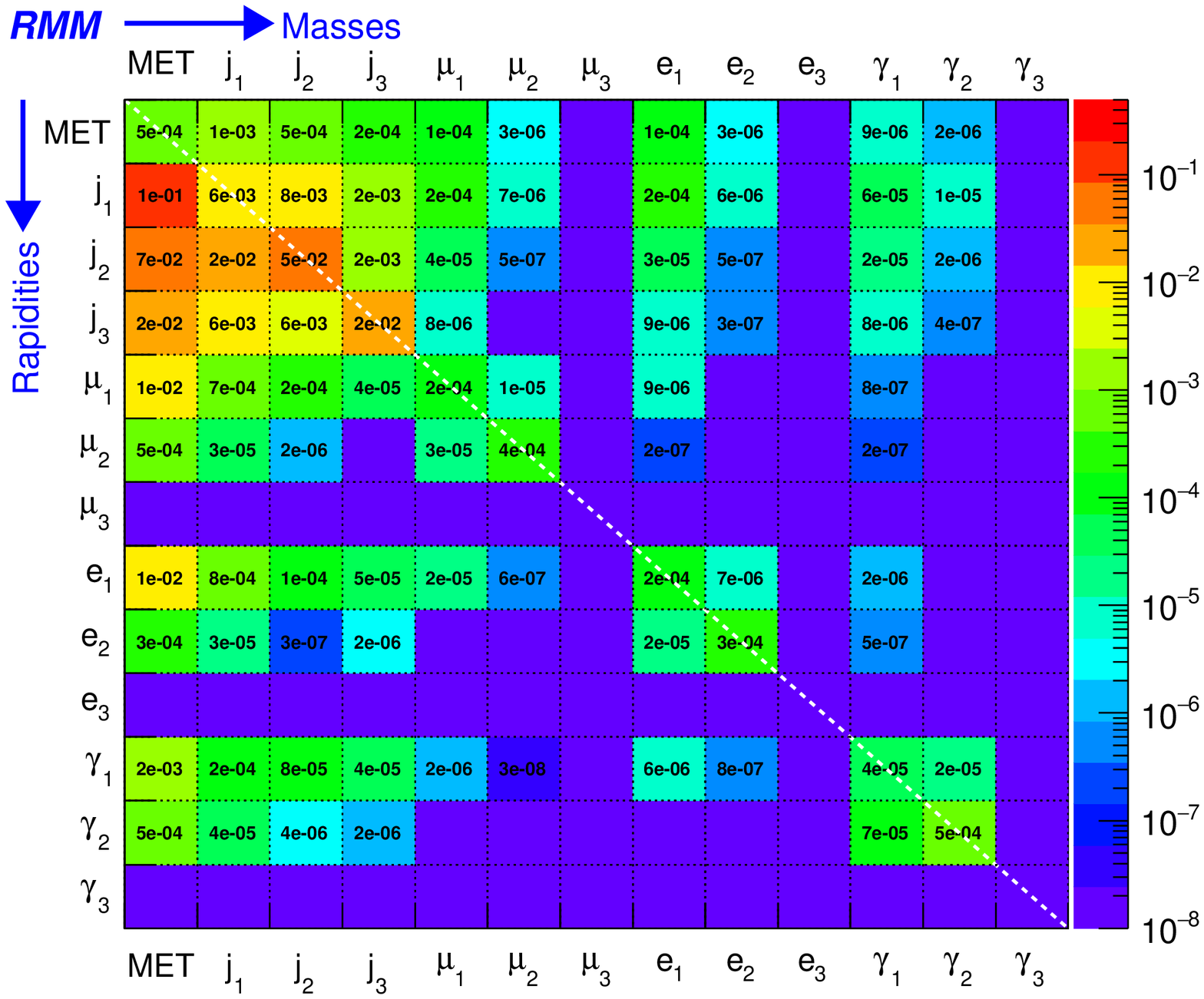}
   }
   \subfigure[Top ($t\bar{t}$) production] {
   \includegraphics[width=0.4\textwidth]{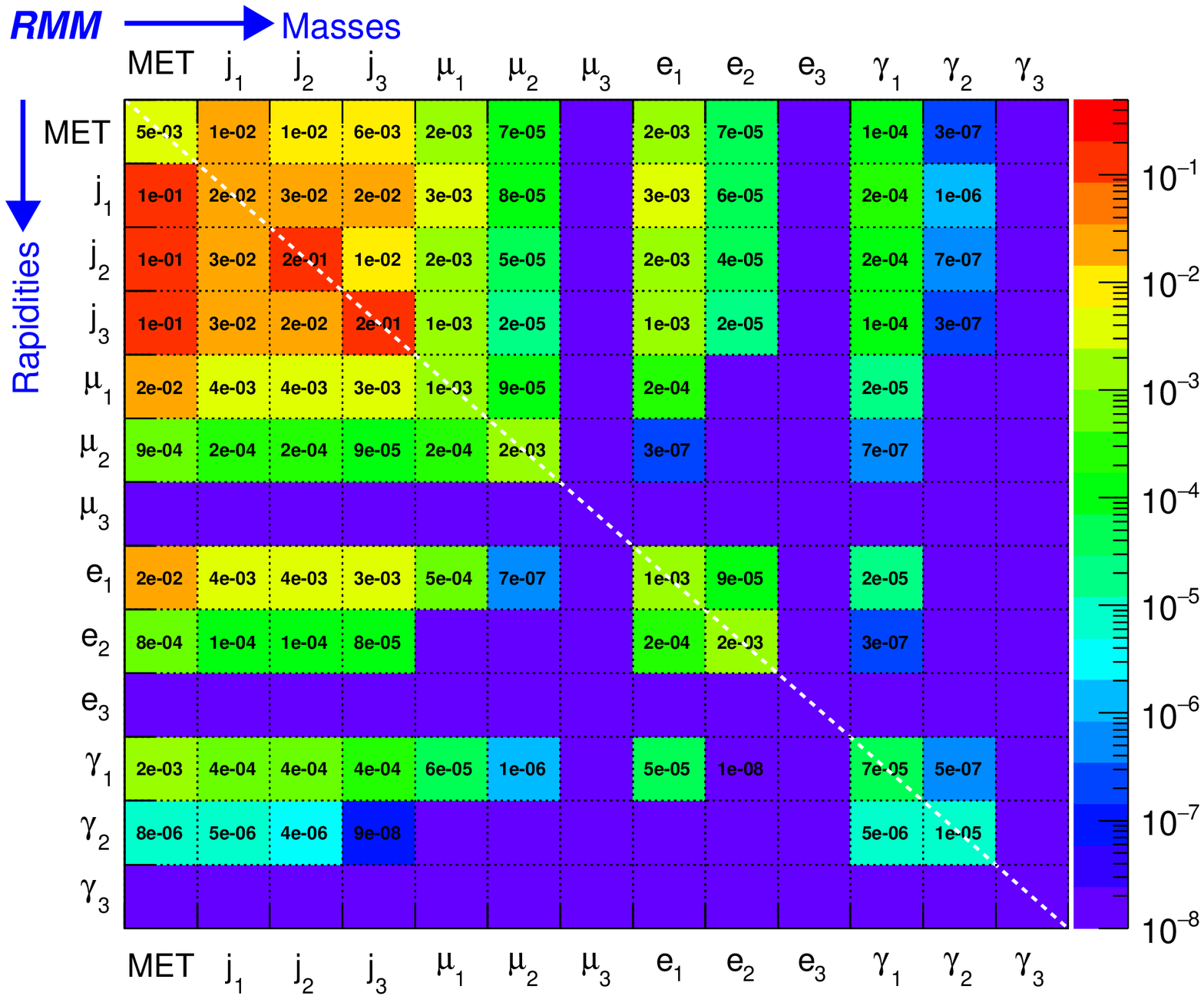}\hfill
   }
   \subfigure[$H^+t$ production] {
   \includegraphics[width=0.4\textwidth]{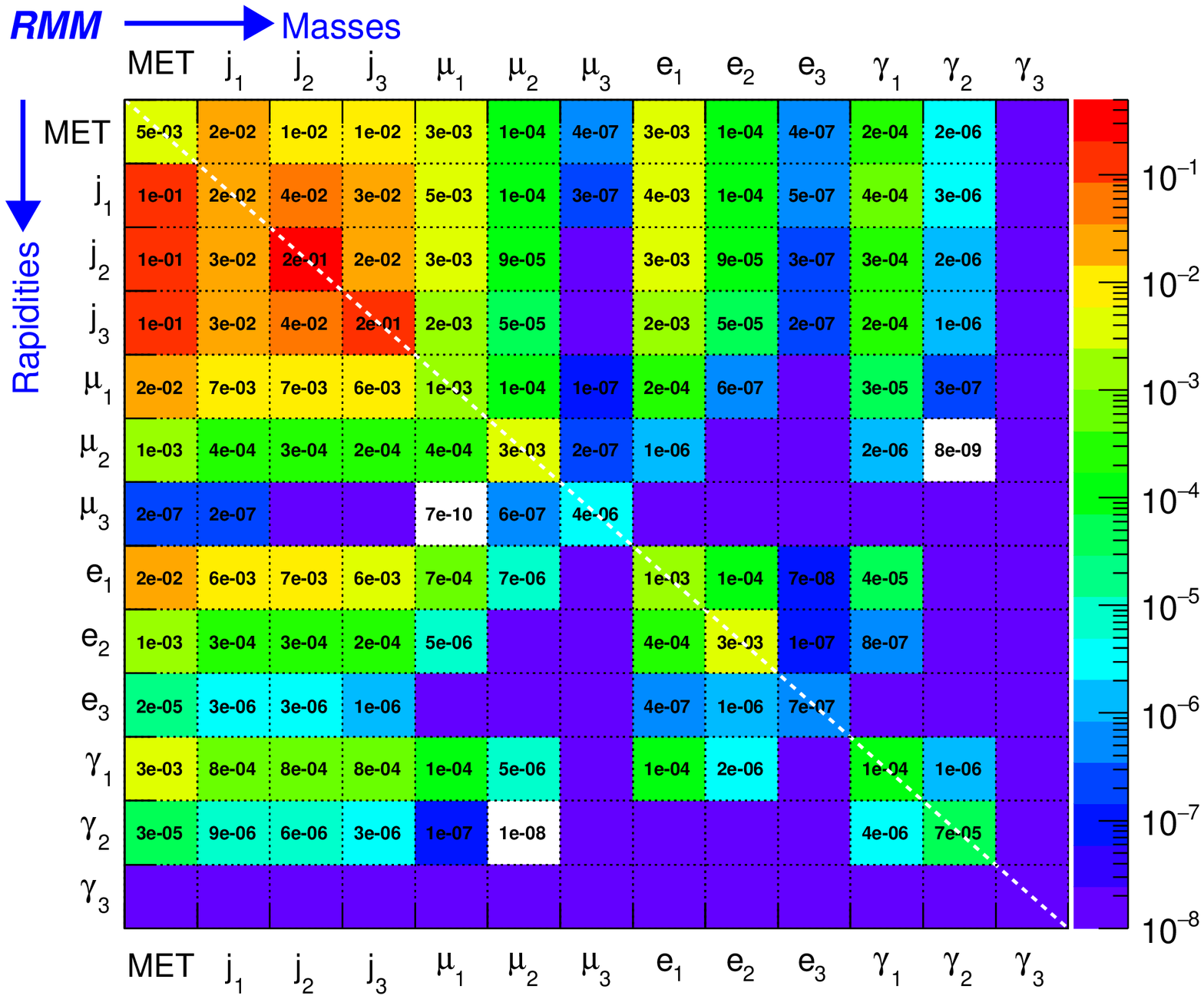}
   }

\end{center}
\caption{Visualization of the T4N3 RMMs for four different processes using 50,000 Monte Carlo
events for each process (with all decay modes allowed for top quarks, vector bosons and $H$).
In the case of  $H^+$ decaying to $W^+$ and $H$, $H$ decays to $b-$jets.
The cell position (1,1) contains the information with a missing transverse energy $\etmiss$  scaled by $1/\sqrt{s}$.
The first row contains  scaled transverse masses $m_T(i)$. 
The white dashed line shows the diagonal elements with transverse
energy and their imbalances. The upper-right part of this matrix contains $m(i,j)$, while the lower-left
part contains rapidity differences, $h(i,j)$.}
\label{fig:prof}
\end{figure}

The software and Monte Carlo settings were taken from the HepSim project \cite{Chekanov:2014fga}.  
Stable particles 
with a lifetime larger than $3\cdot 10^{-10}$ seconds were considered, while 
neutrinos were excluded from consideration.
The jets were
reconstructed with the anti-$k_T$ algorithm \cite{Cacciari:2008gp} as implemented in the {\sc FastJet} package~\cite{Cacciari:2011ma} using a distance parameter of $R=0.6$. 
The minimum transverse energy of jets was $50$~GeV, and the pseudorapidity range was $|\eta|<3$.
Muons, electrons and photons were reconstructed from the truth-level record, making sure that they are
isolated from jets. The minimum transverse momentum of leptons and photons was 25~GeV. 
The missing transverse energy
is recorded only if it is above 50~GeV. 

For a better understanding the RMM formalism, Fig.~\ref{fig:event} shows two events with $t\bar{t}$ from the Pythia8 generator.
Fig.~\ref{fig:event}(a) shows an event where $t$ and $\bar{t}$ quarks decay
to six jets, while  Fig.~\ref{fig:event}(b) shows an event where one top quark 
decays to $\bar{b}W^+$ with $W^+\to e^+\nu_e$.
The figures show rather distinct patterns for these two events.
The six-jet event does not have leptons and $\etmiss$  (at the position (1,1)).
Note that the simple T4N3 configuration used in this paper to illustrate the RMM concept is not appropriate 
for real-life cases since it cannot accommodate all reconstructed jets, nor $b-$jets. 
How to determine the RMM configuration will be explained at the end of Sect.~\ref{sec:ev1}.

The RMMs matrices can also be used for visualizing  groups with many events.
Figure~\ref{fig:prof} shows the average values of T4N3 RMM cells
for the Monte Carlo events described above. 
As in the case of single events, the figures show  rather distinct patterns for these four processes.
The QCD multijet events do not have large values of $\etmiss$  at the position (1,1), and there is no 
large rates of leptons and photons. 
The Higgs processes have an enhanced production of two photons and leptons, indicated with significant values at ($\gamma_1, \gamma_2$),
($\mu_1, \mu_2$) and ($e_1, e_2$).
If one considers ($\gamma_1, \gamma_2$) cells only,
the Higgs mass can be reconstructed  from
the $m(\gamma_1, \gamma_2)$ cells shown in Fig.~\ref{fig:prof}(b),
after multiplying their values
by $\sqrt{s}$.
The $t\bar{t}$  events have  a large missing transverse energy at (1,1)  and a significant jet activity.  
The largest similarity between RMMs was found  for the $H^+$ and $t\bar{t}$ events shown in 
Figs.~\ref{fig:prof}(c) and (d). 
If a single decay channel is considered for $H^+ \to W^+H$, such
as $H$ decaying to two photons, the RMM for $H^+$ will be significantly different 
from the other processes.

\section{RMM for Neural Networks}

\begin{figure}[ht]
\begin{center}
   \subfigure[] {
   \includegraphics[width=0.4\textwidth]{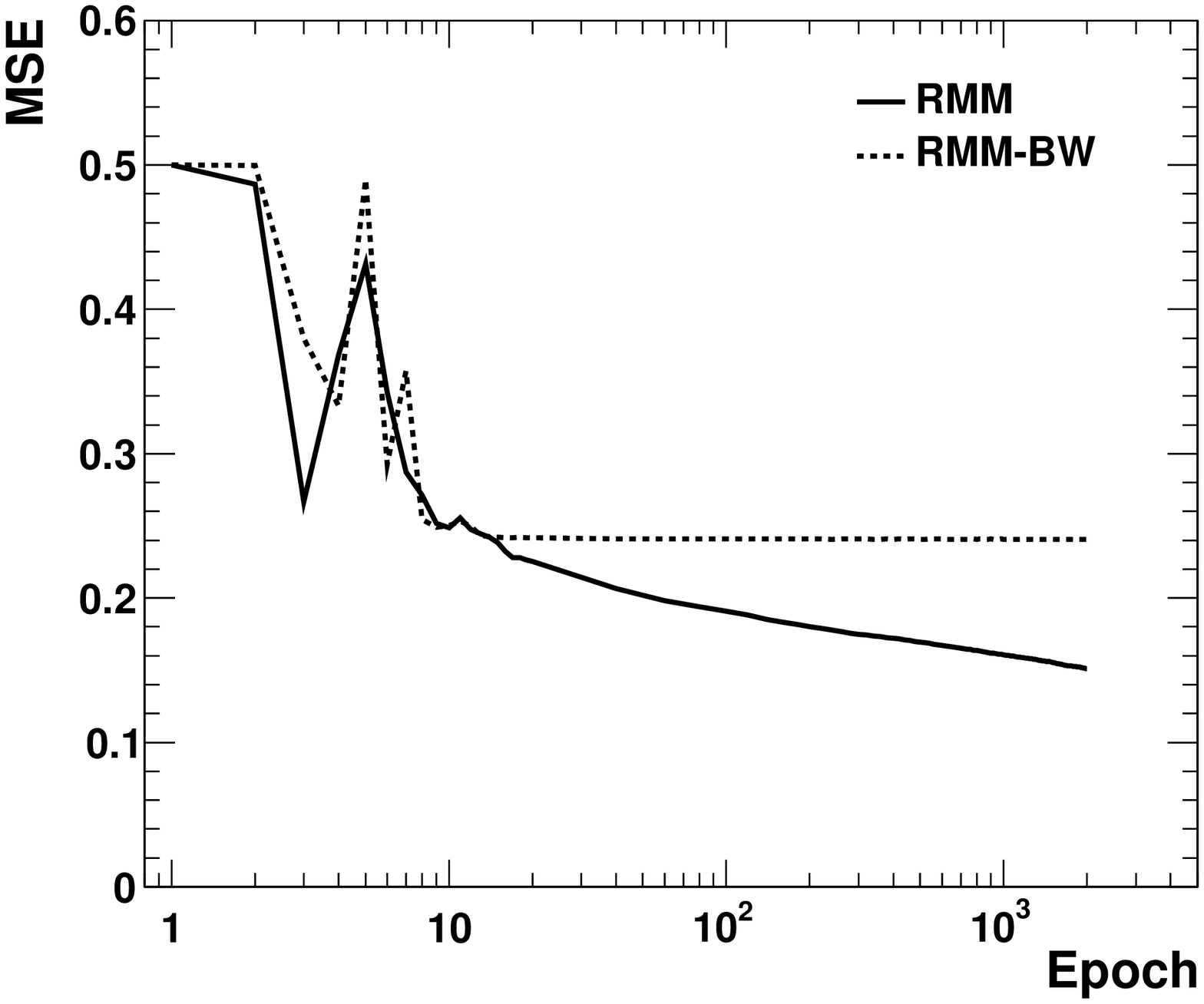}\hfill
   }
   \subfigure[] {
   \includegraphics[width=0.4\textwidth]{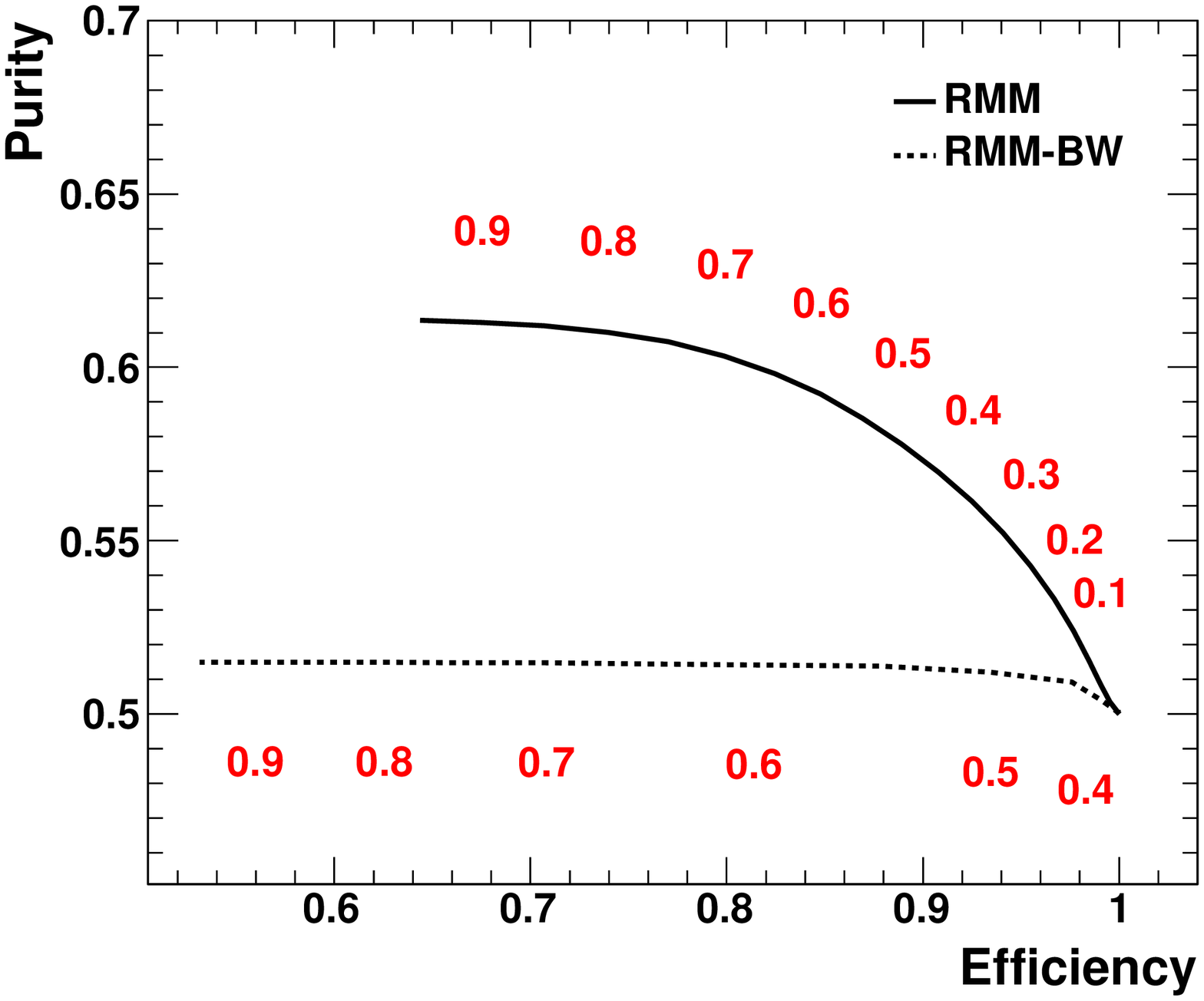}
   }
\end{center}
\caption{(a) MSE values versus epoch for the separation of $H^+$ 
from background $t\bar{t}$ events using the NN. The solid line shows the standard RMM, while
the dashed  line shows a version  
of the RMM where non-zero values are replaced by 1 (RMM-BW).
A shallow backpropagation NN was used with 169 input nodes, 120 nodes in a single hidden layer,
and one node in the output layer.
10,000 events of each category were used for the NN training. The training was terminated using a cross validation sample.
Figure (b) shows the reconstruction efficiency of $H^+$ events versus the reconstruction purity
for the standard RMM and RMM-BW. The values of the output node at fixed values
of the efficiency are shown above (below) the curves. i
}
\label{fig:eff}
\end{figure}

An identification of the Standard Model processes, such as multijet QCD, Higgs and $t\bar{t}$ does not represent 
any challenge for the NN classification, since even the visualized RMMs show different patterns for these processes.
However, the separation of $H^+$ events from $t\bar{t}$ is more difficult due to the similarity of their RMMs. 
The  main  distinct feature of these two processes
is the RMM cell values, i.e. the color patterns shown in Fig.~\ref{fig:prof}, not the numbers of non-zero cells in events.
Therefore, the $H^+$ and $t\bar{t}$ events were used to verify the classification capabilities of the RMM. 

As  simple test of the RMM concept for event classification using machine learning,
we have created 10,000 RMMs for $H^+$ and  $t\bar{t}$ events, which are then  
used as the input for a shallow backpropagation NN with the sigmoid activation function.
The NN was implemented using the {\sc FANN} package \cite{fann}.
No re-scaling of the input values was applied since the value range of $[0, 1]$ 
is fixed by the definition of the RMM.
The NN had 169 input nodes, which were mapped to a 1D array obtained from  
T4N3 RMM with a size of $13\times 13$. A single hidden layer had 120 nodes, while 
the output layer had a single node.
During the NN training,  the output node value 
was set to 0 for $t\bar{t}$ events and to 1 for $H^+$ events.  
This value corresponds to a  probability that
a given collision event belongs to the $H^+$ category.
The NN was trained using 2,000 epochs. The training was terminated using an independent (''cross-validation'') 
sample with 10,000  RMMs. It was found that  
the MSE of the validation sample does not decrease with 
the number of epochs after 2,000 epochs.   

To understand the performance of the NN, the default activation function was changed to
a linear activation function  in the {\sc FANN} package \cite{fann}. 
The number of nodes in the hidden layer was varied in the range (50-400).  
In addition, the number of hidden nodes was increased to two. 
No significant changes
in the NN performance were found.

Figure~\ref{fig:eff}(a) shows that  the   
mean square error (MSE) of the NN based on the RMM decreases as a function of the epoch number (the solid line).
This indicates a well-behaved NN training.
A spike for small number of epochs is due the gradient-descent optimization     
applied  to the  sparse input data. 
In addition to the standard RMM, 
this figure shows a special case when all values of the  RMM were  set to 1 for cells with non-zero
values (i.e. converting the RMMs to ``black-and-white'' images).   
Such 2D arrays, called RMM-BW, were  constructed to check the  sensitivity of the NN output to the amplitude of 
the cell values. 
According to Fig.~\ref{fig:eff}(a), the NN with the RMM-BW inputs cannot effectively be trained, since
the MSE values are independent of the epoch number. 
Therefore, for the given example, 
the number of cells containing zero values in the RMM (and thus the multiplicities of objects) 
is not an important factor for the NN training.

To verify the performance of the trained NN, a sample of RMMs was constructed 
from $t\bar{t}$ and $H^+$ events
which were not used during the training procedure. 
Then the trained NN was applied for predicting the output node value.
The success of the NN training was evaluated in terms of the purity of the reconstructed 
$H^+$ events as a function of the reconstructed efficiency. 
The efficiency of identification of $H^+$ events was defined as 
a fraction of the number of true $H^+$ events 
with the NN output above some value.
This value can be varied between  0 and 1, 
with 1 being the most probable likelihood that the event belongs to $H^+$ process.
We also calculated the purity of the reconstructed $H^+$ events as a ratio of the number of $H^+$ events that 
met a requirement on the NN output value,  
divided by the number of accepted events (irrespective of the origin of these events).
Both the efficiency and purity depend on the value of the output node. 

Figure~\ref{fig:eff}(b) shows the performance of the NN event classifier in terms of 
the efficiency to identify the $H^+$  events versus the event purity for several requirements on the NN output value. 
For the used Monte Carlo event samples, the statistical uncertainties on the purity-efficiency curves 
are less than 3\%.

According to this figure, the standard RMM can be used to identify $H^+$ events. 
The RMM-BW input for the NN  fails for the event 
classification due to the lack of influential features.

\section{Event classification for searches}
\label{sec:ev1}


\begin{figure}[ht]
\begin{center}
   \subfigure[] {
   \includegraphics[width=0.4\textwidth]{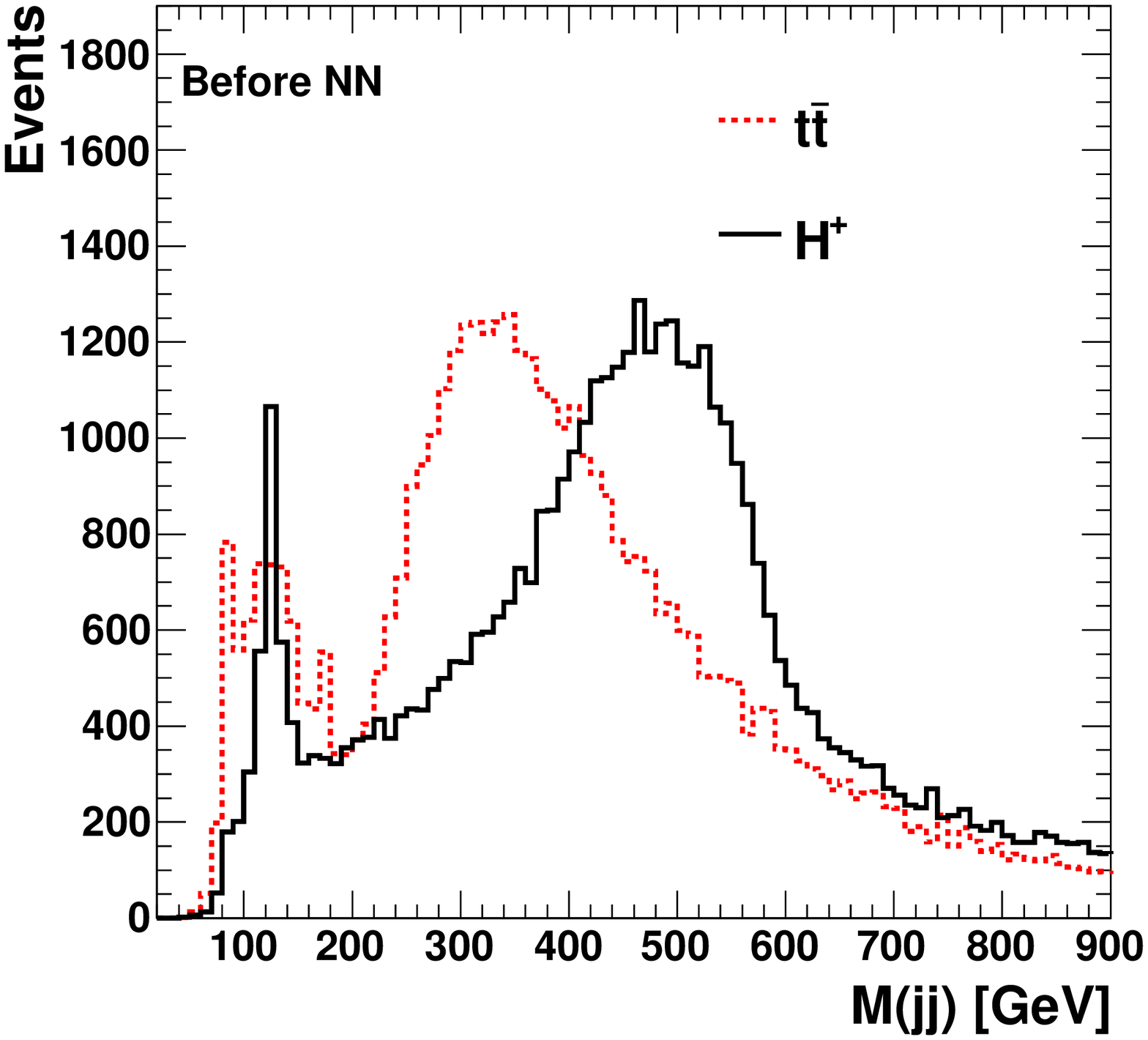}\hfill
   }
   \subfigure[] {
   \includegraphics[width=0.4\textwidth]{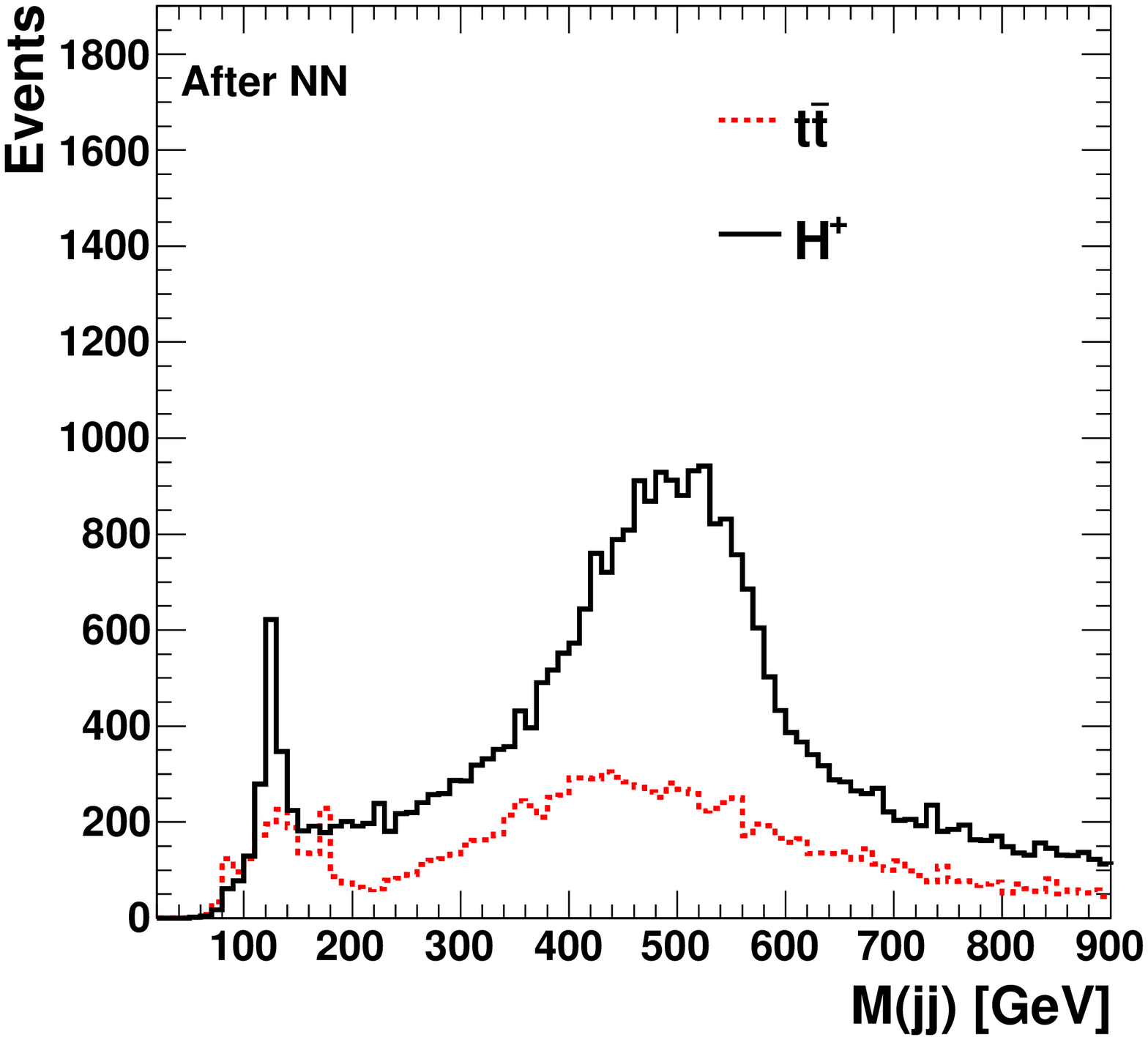}
   }
\end{center}
\caption{Invariant masses of two leading in $E_T$ jets  for $t\bar{t}$ and $H^+$ event categories before (a) and after (b) applying the NN
based on RMM inputs and requiring the NN output values above 0.5.
The NN was trained after disabling the cells that carry information on the masses of two leading jets.
The number of events for each event category ($t\bar{t}$ and $H^+$) is 50,000. 
The dijet masses were reconstructed from the RMM cell (3,2).
Statistical uncertainties for each histogram are given by the square root of the number of entries in each bin (not shown).
}
\label{fig:mass}
\end{figure}

The RMMs  can project complex collision events into to a fixed-size phase space, 
such that the mapping  of the RMM cells  to the NN with a fixed number of the input nodes is unique, independent of
how many objects are produced.
A well-defined input space is the requirement for many machine learning algorithms. 
At the same time, the RMM covers almost every aspect
of particle/jet production in terms of major (weakly correlated) kinematic variables.
Therefore, studies of the most influential  
variables used for classification of events can be reduced or even avoided. 
However, in some situations, 
in order to reduce biases that can distort
background distributions towards signal features,
RMM cells with variables to be used for physics measurements need to be 
identified and disregarded during  machine learning.

Let us consider a simple example. We will separate the $H^+$ events (``signal'') from $t\bar{t}$ (``background'') using
the NN based on the RMMs. 
As a measure of our success in the event classification, we will use the 
invariant masses of two leading in $E_T$ jets which correspond 
to the cell position (3,2). This variable will also be used to perform physics studies, such as   
the measurement of the $H^+$ mass. 
The distribution of this variable should (preferably) be unbiased  by the NN training. 
Therefore, we will disable $e_T(i)$ and $m(i,j)$ inputs  
at the RMM positions (2,2) and (3,2) during the NN training.

Figure~\ref{fig:mass}(a)
shows the dijet mass distribution reconstructed from the RMM after summing up the 
cell values at the position (3,2) over all events, and scaling the sum by $\sqrt{s}$. 
We use 50,000 events for each process category, i.e. for the signal ($H^+$) and background events ($t\bar{t}$).  
A sharp peak observed at 125~GeV is due to the decay $H\to b\bar{b}$, when each $b$-quark forms a jet. 
A broad peak
near $500$~GeV is due to the decays of $H^+$ to $W(\to q\bar{q})$ and $H(\to b\bar{b})$, when the decay
products of $W$ and $H$ are merged by the jet algorithm to form two jets. 
The peak position of the latter is shifted from the generated $H^+$ mass
of 600~GeV. This is due to the fact that the jet algorithm has a size of 0.6, 
which is not large enough to collect all hadronic activity
from boosted $W$ and $H$.
The $t\bar{t}$ events generated as explained above represent a background 
which masks the $H^+$ signal. 
Note that, in real-life scenarios, the rate of the background $t\bar{t}$ events is significantly
larger than that given in this toy example.

The NN with the two disabled input links described above was trained using 10,000 events with $t\bar{t}$ and $H^+$.
The trained NN was applied to classify the event sample with 50,000 events per each event category. 
We accept $H^+$ events when the value
of the output node is above 0.5. After applying the NN selection, one can successfully reduce the $t\bar{t}$ background, as shown in Fig.~\ref{fig:mass}(b).
The background distribution is somewhat shifted towards the signal peak, indicating that
some residual effect from the NN training is still present. 
But this bias cannot change our main conclusion that the signal can reliably be  
identified.
The NN selection decreases the background at the  500~GeV mass region 
by a factor 3,  while the $H^+$ signal rate is only reduced by $30\%$.
Note that the background contribution to the $H^+$ events 
depends on the realistic event rates of the $t\bar{t}$ and $H^+$ processes, which are not considered in this example.

A comment on evaluation of systematic uncertainties when using the RMM should follow. 
As in the case of the usual cut-and-reject methods,  the  RMM matrices should be recalculated 
using different conditions for all reconstructed objects.
In the case of calculations of statistical limits, the histograms obtained after applying the RMM selections 
should be included in dedicated programs for limit calculations.

We believe that the  event classification capabilities of this approach can significantly be increased
by considering RMMs with more than three jets, leptons with different charges, 
$b$-jets, reconstructed $\tau$ leptons, or using multi-dimensional matrices
with three or more particle correlations.
To improve classification, a care should be taken to avoid a ``saturation'' effect  when
multiplicities of particles/jets are larger than the chosen $N_i$ that define the RMM size, otherwise 
the loss of useful information may prevent an effective usage of the RMM.
We used $N_i=3$ in the simple example described above for better explanation and visualization of this method.
This is not a sufficiently large number for the chosen transverse momentum cut in order to
accommodate  events with a large number of jets.
Note that larger values of $N_i$ do not increase 
dramatically the sizes of the files used for storing sparse matrices since only non-zero
values of cells (and their indices) are kept.
A  more sophisticated event classification and comparisons  
with alternative machine learning techniques are beyond  the computational 
resources dedicated to this paper. Other examples of event classifications using the RMM approach
were considered in Ref.~\cite{Chekanov:2018zyv}.

\section{Conclusion}
We propose a method to transform events from colliding experiments to 
the  language widely used by machine learning algorithms, i.e.
fixed-size sparse matrices. In addition, the RMM transformation                   
can be viewed as an effective mapping of complex collision data to pixelated representation useful for visual study.
The method does not exclude the use of different types of
neural network (deep or recurrent) or other machine learning techniques.

By construction, groups of RMMs cells that belong to certain types of objects 
are connected by proximity due to a well-defined hierarchy of the kinematic variables.  
Therefore, the usage of the RMM in particle physics 
may leverage widely used algorithms developed for image identification that
exploit local connectivity of pixels (cells).

Our tests indicate  that the proposed  approach of imaging collision data for event classification
can be useful for preparing a feature space for machine learning. 
The RMM method is sufficiently general  and, typically,  does not require 
detailed studies of influential variables sensitive to background events. But care should be taken  
to avoid using NN inputs that may bias the shapes of observables which will be used later in searches for
new physics. The C++ library that transforms the event records to the RMM is available \cite{map2rmm}. 

\section*{Acknowledgments}

I would like to thank J.~Proudfoot and J.~Adelman for useful discussion.
The submitted manuscript has been created by UChicago Argonne, LLC, Operator of Argonne National Laboratory (“Argonne”). Argonne, a U.S. 
Department of Energy Office of Science laboratory, is operated under Contract No. DE-AC02-06CH11357. The U.S. Government retains for itself, 
and others acting on its behalf, a paid-up nonexclusive, irrevocable worldwide license in said article to reproduce, prepare derivative works, 
distribute copies to the public, and perform publicly and display publicly, by or on behalf of the Government.  
The Department of Energy will provide public access to these results of federally sponsored research in accordance with the 
DOE Public Access Plan. \url{http://energy.gov/downloads/doe-public-access-plan}. Argonne National Laboratory’s work was 
funded by the U.S. Department of Energy, Office of High Energy Physics under contract DE-AC02-06CH11357.


\newpage
\clearpage
\bibliography{biblio}

\clearpage
\appendix
\renewcommand{\thesubsection}{\Alph{subsection}}
\section*{Appendices}
\addcontentsline{toc}{section}{Appendices}

\section{Correlation coefficients of the RMM}
\label{app:corr}

\begin{table}[!htb]
    \caption{RMM cells with the values of the Pearson correlation coefficients ($\rho$) larger than 15\%.
The coefficients were  calculated from the 
Pythia8 Monte Carlo simulations described in Sect.~\ref{sec:vis}.  The first column shows the 
correlation coefficients, 
the second column shows the position of the first cell, and the third column shows the position of the second cell. 
In the RMM notation, 
the largest correlation (92\%) is observed for the transverse masses  calculated  
using leading and sub-leading jets.}

{\tiny  
    \begin{minipage}{.25\linewidth}
      \centering
        \begin{tabular}{c|c|c}
$\rho$  &  (r1,c1)  & (r2, c2) \\ \hline  
0.92 &  2,1 & 3,1 \\
0.87 &  5,2 & 5,5 \\
0.87 & 1,1 & 2,1 \\
0.83 &  1,1 & 3,1 \\
0.82 &  3,1 & 4,1 \\
0.81 &  5,2 & 5,3 \\
0.80 &  6,2 & 6,3 \\
0.80 &  6,1 & 6,2 \\
0.79 &  6,2 & 6,5 \\
0.79 &  5,3 & 5,5 \\
0.79 &  4,6 & 6,4 \\
0.79 &  2,2 & 3,2 \\
0.78 &  2,1 & 4,1 \\
0.77 &  2,6 & 6,2 \\
0.76 &  6,1 & 6,5 \\
0.76 &  3,6 & 6,3 \\
0.75 &  5,3 & 5,4 \\
0.74 &  6,3 & 6,5 \\
0.74 &  5,1 & 5,2 \\
0.74 &  4,5 & 5,4 \\
0.74 &  1,1 & 4,1 \\
0.72 &  6,5 & 6,6 \\
0.72 & 5,6 & 6,5 \\
0.71 &  5,2 & 5,4 \\
0.70 &  3,5 & 5,3 \\
0.69 &  5,1 & 5,5 \\
0.69 &  1,6 & 6,1 \\
0.68 &  6,3 & 6,4 \\
0.68 &  6,1 & 6,3 \\
0.67 &  5,4 & 5,5 \\
0.67 &  3,4 & 4,3 \\
0.66 &  6,1 & 6,6 \\
0.65 &  5,1 & 5,3 \\
0.65 &  1,6 & 6,2 \\
0.63 &  6,2 & 6,6 \\
0.63 &  5,6 & 6,2 \\
0.62 &  2,5 & 5,2 \\
0.62 &  1,6 & 2,6 \\
0.61 &  3,6 & 6,4 \\
0.60 &  4,2 & 4,3 \\
0.60 &  3,6 & 4,6 \\
0.59 &  2,6 & 5,6 \\
0.59 &  2,4 & 4,2 \\
0.58 &  6,2 & 6,4 \\
0.58 &  2,2 & 4,2 \\
0.57 &  3,2 & 4,2 \\
0.57 &  2,6 & 6,1 \\
0.55 &  5,6 & 6,3 \\
0.54 &  6,3 & 6,6 \\
0.54 &  5,1 & 5,4 \\
0.54 &  3,6 & 6,2 \\
0.53 & 5,6 & 6,1 \\
\end{tabular}

    \end{minipage}%
    \begin{minipage}{.25\linewidth}
      \centering
        \begin{tabular}{c|c|c}
$\rho$ &  (r1,c1)  & (r2, c2) \\ \hline  
0.53 &  1,6 & 6,5 \\
0.53 &  1,6 & 6,3 \\
0.53 &  1,6 & 5,6 \\
0.52 &  1,5 & 2,5 \\
0.51 &  2,6 & 6,5 \\
0.51 &  2,6 & 6,3 \\
0.50 &  6,4 & 6,5 \\
0.49 &  3,6 & 6,5 \\
0.49 &  3,2 & 4,3 \\
0.48 &  4,6 & 6,3 \\
0.47 & 3,6 & 5,6 \\
0.47 &  1,6 & 3,6 \\
0.47 &  1,5 & 3,5 \\
0.46 &  1,5 & 5,3 \\
0.46 &  1,5 & 5,2 \\
0.46 &  1,5 & 5,1 \\
0.45 &  3,6 & 6,1 \\
0.45 &  2,5 & 5,3 \\
0.45 & 2,5 & 3,5 \\
0.45 &  1,5 & 4,5 \\
0.44 &  6,1 & 6,4 \\
0.44 &  3,5 & 5,4 \\
0.44 &  3,5 & 5,2 \\
0.44 &  2,6 & 3,6 \\
0.43 &  5,6 & 6,6 \\
0.43 &  4,5 & 5,3 \\
0.43 &  2,6 & 6,6 \\
0.43 &  1,6 & 6,6 \\
0.43 &  1,5 & 5,4 \\
0.42 & 5,6 & 6,4 \\
0.42 &  3,5 & 4,5 \\
0.42 &  3,2 & 4,4 \\
0.41 &  2,3 & 3,2 \\
0.40 &  2,5 & 5,4 \\
0.40 &  2,5 & 4,5 \\
0.40 &  2,2 & 4,3 \\
0.40 &  1,4 & 2,4 \\
0.39 &  4,5 & 5,2 \\
0.38 &  2,5 & 5,1 \\
0.37 &  1,6 & 6,4 \\
0.37 &  1,6 & 4,6 \\
0.37 &  1,4 & 3,4 \\
0.37 &  1,1 & 5,1 \\
0.36 &  6,4 & 6,6 \\
0.36 &  3,5 & 5,1 \\
0.35 &  3,6 & 6,6 \\
0.34 & 3,5 & 5,5 \\
0.34 & 2,2 & 4,4 \\
0.34 &  1,5 & 5,5 \\
0.33 & 4,6 & 6,2 \\
0.33 &  2,4 & 3,4 \\
0.32 &  2,6 & 6,4 \\
\end{tabular}

    \end{minipage} 
     \begin{minipage}{.25\linewidth}
      \centering
        \begin{tabular}{c|c|c}
$\rho$ &  (r1,c1)  & (r2, c2) \\ \hline  
0.29 &  4,5 & 5,5 \\
0.29 &  3,1 & 5,1 \\
0.29 &  2,1 & 2,2 \\
0.29 &  1,4 & 4,3 \\
0.29 &  1,4 & 4,2 \\
0.28 &  4,6 & 6,1 \\
0.28 &  2,1 & 3,3 \\
0.27 &  4,6 & 6,5 \\
0.27 &  4,6 & 5,6 \\
0.27 &  2,4 & 4,3 \\
0.26 &  4,1 & 4,3 \\
0.26 &  2,3 & 3,4 \\
0.25 &  2,2 & 3,3 \\
0.25 &  1,3 & 3,4 \\
0.25 &  1,2 & 2,3 \\
0.24 &  4,3 & 4,4 \\
0.24 &  4,2 & 4,4 \\
0.24 & 3,4 & 4,2 \\
0.24 &  1,1 & 3,3 \\
0.23 &  4,1 & 5,1 \\
0.23 &  3,3 & 4,4 \\
0.23 &  2,2 & 3,1 \\
0.22 &  5,5 & 6,6 \\
0.22 &  4,1 & 4,2 \\
0.22 &  2,3 & 2,4 \\
0.22 &  1,2 & 2,4 \\
0.21 &  5,5 & 6,5 \\
0.21 &  5,1 & 6,1 \\
0.21 &  4,6 & 6,6 \\
0.21 &  2,1 & 5,2 \\
0.20 &  4,1 & 5,4 \\
0.18 &  5,1 & 6,6 \\
0.18 &  3,1 & 5,3 \\
0.18 &  1,4 & 4,1 \\
0.17 &  5,5 & 6,2 \\
0.17 &  5,5 & 6,1 \\
0.17 &  5,1 & 6,5 \\
0.17 &  5,1 & 6,2 \\
0.17 &  2,6 & 4,6 \\
0.17 &  1,1 & 5,2 \\
0.16 & 3,3 & 4,3 \\
0.16 &  3,1 & 3,3 \\
0.16 & 3,1 & 3,2 \\
0.16 &  2,2 & 4,1 \\
0.16 &  2,1 & 5,3 \\
0.15 &  5,5 & 6,3 \\
0.15 &  5,2 & 6,5 \\
0.15 &  5,2 & 6,2 \\
0.15 &  3,4 & 4,1 \\
0.15 &  3,1 & 5,2 \\
0.15 &  1,4 & 4,4 \\
\end{tabular}

    \end{minipage}
}
\end{table}

\end{document}